\documentclass[12pt,preprint]{aastex}

\newcommand{\etal}{et\,al.}
\newcommand{\av}{$A_V$}
\newcommand{\ejk}{$E(J-K_S)$}
\newcommand{\tausil}{$\tau_{9.7}$}

\begin{document}

\title{The Relationship between the Optical Depth of the 9.7 \micron\ Silicate
Absorption Feature and Infrared Differential Extinction in Dense Clouds}

\author{J.E. Chiar\altaffilmark{1} and
      K. Ennico\altaffilmark{2} and
      Y. J. Pendleton\altaffilmark{3} and
      A.C.A. Boogert\altaffilmark{4} and
      T. Greene\altaffilmark{2} and
      C. Knez\altaffilmark{5} and
      C. Lada\altaffilmark{6} and
      T. Roellig\altaffilmark{2} and
      A. G. G. M. Tielens\altaffilmark{3} and
      M. Werner\altaffilmark{7} and
      D.C.B. Whittet\altaffilmark{8}}
      
\altaffiltext{1}{SETI Institute, 515 N. Whisman Road, Mountain View, CA 94043} 
\altaffiltext{2}{NASA Ames Research Center, Mail Stop 245-6, Moffett Field, CA
94035}
\altaffiltext{3}{NASA Ames Research Center, Mail Stop 245-1, Moffett Field, CA
94035}
\altaffiltext{4}{AURA/NOAO-South, Casilla 603, La Serena, Chile}
\altaffiltext{5}{University of Texas, 1 University Station, Austin, TX 78712}
\altaffiltext{6}{SAO, 60 Garden Street, Cambridge, MA 02138}
\altaffiltext{7}{JPL, 4800 Oak Grove Drive, Pasadena, CA 91109}
\altaffiltext{8}{Rensselaer Polytechnic Institute, Troy, NY 12180}

\begin{abstract}

We have examined the relationship between the optical depth of the 9.7
\micron\ silicate absorption feature (\tausil) and the near-infrared
color excess, \ejk, in the Serpens, Taurus, IC 5146,  Chameleon I, Barnard 59, and Barnard 68 dense clouds/cores. Our data set, based largely on Spitzer IRS spectra,
spans \ejk=0.3 to 10 mag (corresponding to visual extinction between about 2
and 60 mag.). All lines of sight show the 9.7 \micron\ silicate feature. 
Unlike in the diffuse ISM where a tight linear correlation between the 9.7
\micron\ silicate feature optical depth and the extinction ($A_V$) is observed,
we find that the silicate feature in dense clouds does not show a monotonic
increase with extinction.  Thus, in dense clouds, $\tau_{9.7}$ is not a good measure of total dust column density.  With few exceptions, the measured  \tausil\ values
fall well below the diffuse ISM correlation line for \ejk$>2$ mag (\av$>$12 mag).  Grain growth via coagulation is a likely cause of this effect.  \end{abstract}


\section{Introduction}

Observationally, interstellar dust is known to consist mainly of amorphous silicates Ñ as evidenced by the 9.7 and 18.5 \micron\ Si-O stretching and bending vibrations of this material, and graphitic carbon, responsible for the strong 2175 \AA\ bump. Detailed models have been developed that link the measured optical properties of these materials to the observed interstellar extinction using appropriate size distributions \nocite{draine_lee84,desert_etal90,zubko_dwek_arendt2004} (e.g., Draine \& Lee 1984; Desert, Boulanger, \& Puget 1990; Zubko, Dwek \& Arendt 2004).   These models differ in their assumptions about the grain populations, whether silicates and graphite are physically separated or mixed in a composite grain with or without ices. Typically, these models conclude that both materials contribute about equal volumes (per H-atom) to the interstellar dust.  The well-known Draine \& Lee (1984) model illustrates that while silicates dominate the extinction beyond about 8 \micron\ through their strong resonances, in the visual through near-IR, graphite dominates the extinction.  In contrast, some composite grain models argue that ``big grains'' made of silicates with a carbon-containing coating (Desert \etal\ 1990) dominate the extinction in both the visible and IR.   Zubko \etal\ (2004) consider a variety of dust components in their models including composite particles (containing silicate, organic refractory material, water-ice, and voids), polycyclic aromatic hydrocarbons (PAHs), bare silicate, graphite and amorphous carbon particles.   The relative contributions of each of these components in the visible, near-IR and mid-IR varies and depends on the specific combination of components considered.  

In the diffuse ISM, observations spanning visual extinction (\av) between 3 and 15 mag have shown that the optical depth of 9.7 \micron\ silicate absorption feature (\tausil) displays a tight linear correlation with \av\ \nocite{roche_aitken84,whittet_bk03} (Roche \& Aitken 1984; Whittet 2003).  This indicates that the silicate and graphite dust components are well-mixed and vary little in relative abundance. In dense clouds, however, there is observational evidence that this correlation suffers a significant break-down. The first indication that denser regions do not follow the diffuse ISM correlation came from the study of \nocite{whittet_etal88} Whittet \etal\ (1988), who found that the correlation appeared to fail at high \av. In that study, two (out of 5) lines of sight in the Taurus dark cloud showed \tausil\ values that were anomalously low with respect to their \av.  

In this paper, we confirm and extend the failing of the \tausil\ versus extinction correlation for lines of sight in the IC 5146, Barnard 59, Barnard 68, Chameleon I, Taurus, and Serpens dense clouds.  We show that a dramatic break-down of the diffuse ISM trend occurs when \av\ exceeds $\sim12$ mag.  Our results are based primarily on Spitzer IRS spectra which we describe in \S\ref{dataredux}.   In \S\ref{correlation} we discuss the relationship between the depth of the 9.7 \micron\ silicate absorption feature, the visual extinction, and near-infrared color excess.  Finally, in \S\ref{discussion}, we discuss the astrophysical implications of these observations in terms of grain properties in dense clouds.  

\section{Data Reduction and Analysis\label{dataredux}}

A summary of our dense cloud dataset including 2MASS IDs, spectral types, silicate optical depth at 9.7 \micron,  $J-K_S$, \ejk, and \facility{Spitzer} Program ID (PID) numbers, where relevant, are given in Table~1.  The observations of the background sources behind IC~5146, Barnard 68 and Chameleon I were part of the Spitzer Cycle 1 program, PID 3320 (PI: Y.J. Pendleton).  The sources were observed using the {\em Infrared Spectrometer\/} (IRS) modules Short-Low (SL) 2 (5.2--8.7 \micron), SL1 (7.4--14.5 \micron), and Long Low (LL) 2 (14.0-21.3 \micron) providing resolving powers ($\lambda/\Delta\lambda$) of 60-127 (SL) and 57-126 (LL2).  These data were processed with the Spitzer pipeline version S12.0 to produce Basic Calibrated Data (BCD).  Post-BCD 1-D spectra were created by carrying out background subtraction and custom extraction in IDL/SMART.   The observations of the Barnard 59 sources were part of the Spitzer Cycle 2 program, PID 20604 (PI: A. Boogert; T.\ L.\ Huard \etal, in preparation).  These data were processed with Spitzer pipeline version 15.3.  The observations of the Serpens sources and the Taurus source, Elias 3, were obtained as part of the the ``c2d'' Legacy program (PI: N. Evans).  The data reduction of these sources is described in Knez et al. (2005; and in preparation).  Silicate optical depths for the Taurus sources Elias 9 and 13 are from Whittet et al. (1988).  The silicate optical depth for Elias 16 is from Bowey \etal\ (1998).  We used the Spitzer IRS spectrum for Elias 3 as it has higher S/N than the spectrum  presented by Whittet et al. (1988).  Spectral types for all Taurus sources are from Whittet \etal\ (1988 and references therein).  For the diffuse ISM, we plot data available in the literature (see \S3) and one additional measurement toward 2MASS 1743320-2847525 (\tausil=1.5, \ejk=4.27; J. Chiar, unpublished Spitzer data, PID 3616).  

To determine the optical depth of the 9.7 \micron\ silicate absorption feature, we fitted a second or third degree polynomial across the combined IRS SL and LL2 (when available) spectra.  The 5.2--7 and 13.5--15 \micron\ regions were chosen to represent the absorption-free continuum.  In cases where LL2 data were available and the 18.5 \micron\ silicate feature was very weak, the region between 20 and 21.3 \micron\ was also used as continuum. The optical depths were then calculated by \tausil\ $= -\ln F_{\rm source}/F_{\rm continuum}$, where $F$ is the flux density in Jansky units.

The $J-K_S$ colors are from the Two-Micron All-Sky Survey (2MASS), except for 2MASS 17112005-2727131, 17111538-2727144 (Barnard 59), and CK2 (Serpens).  These sources were not detected in the 2MASS $J$ band, and only an upper limit is given in the 2MASS catalog.  For 2MASS 17112005-2727131, the source is  also not detected in the $H$ band, so we used more sensitive ground-based $H$ and $K_S$ photometry from \nocite{romanzuniga_etal2007} Rom{\'a}n-Z{\'u}{\~n}iga \etal\ (2007) to estimate the $J$-band color.  For all three sources, the $J-K_S$ color is estimated by assuming median intrinsic colors appropriate for G0-M4 giants (Bessell \& Brett 1988) and using the color-excess ratio $E(J-H)/E(H-K_S)=1.73$ determined by \nocite{indebetouw_etal05} Indebetouw \etal\ (2005) for lines of sight in the Galactic plane.

All sources in Table~1 have 2MASS colors consistent with reddened background giant stars without intrinsic infrared excesses.  In this way, we include only field stars located behind the dense cloud or core, and young stellar objects are excluded.  In most cases, due to the intervening extinction, the field stars are not observable in the visible, and the visual extinction toward these stars cannot be directly measured.  Thus, the near-IR differential extinction (color excess) is measured, and the visual extinction is inferred from that, based on an assumed extinction law.  To eliminate the uncertainty that might be introduced by assuming a specific extinction law, we plot the near-IR differential extinction, \ejk, in Figure 1.  There is some uncertainty introduced for sources where the spectral types are not precisely known.  This uncertainty is represented by the horizontal error bars in Fig. 1, and is not significant enough to change the trend of the data.  

\section{The Relationship between \ejk\ and \tausil\label{correlation}}

Figure 1 shows a plot of the  measured optical depth of the 9.7 \micron\
silicate absorption feature versus the near-IR differential extinction for the Taurus, 
IC~5146, Chameleon~I, Serpens, Barnard~59 and Barnard~68 dense clouds. 
The solid line is the correlation line, \tausil\ $=A_V/18.0$, calculated based on 13 lines of sight in the diffuse ISM with \av\ between 3 and 29 mag (\S2; Roche \& Aitken 1984; Whittet 2003 and references therein), excluding the Galactic center\footnotemark. \footnotetext{Sources toward the Galactic center fall well {\it above} the correlation line due to their strong silicate absorption features (\tausil$=3.2-4.0$; Roche \& Aitken 1985) relative to their \av\ and color excess \nocite{becklin_etal78IV} ($E(J-K)\sim5.0$; Becklin \etal\ 1978).  There, it is thought that the lack of carbon-rich dust in the central regions of the Galaxy reduces the extinction in the visual and near-IR relative to the extinction in the 10 \micron\ region.} \nocite{whittet_bk03,roche_aitken84}  In order to plot this correlation on the \tausil\ versus \ejk\ plot, we converted $A_V$ to near-IR differential extinction using the diffuse ISM extinction law given by Whittet (2003): $A_V=6.2\times$\ejk.    Comparing the loci of the dense cloud data with the diffuse ISM correlation line, we note the following.  For \ejk$<2$, the dense cloud sources follow the diffuse ISM correlation, showing a monotonic linear increase of \tausil\ with \ejk.  Compared to the diffuse ISM trend, the dense clouds show a markedly weak silicate absorption feature relative to near-IR extinction for \ejk$>2$ mag (\av$\sim12$ mag). With the exception of two lines of sight, \tausil\ reaches a maximum value 0.85 for \ejk\ up to almost 10 mag (\av$\sim60$ mag).  The outliers are two sources in the Serpens dark cloud.  The Serpens source, SSTc2d182852.7+02824, is the only source in our dataset with \ejk$>2$ whose silicate optical depth places it on the diffuse ISM correlation line.  The
highly extincted Serpens field star, CK2, stands out as having a very deep silicate
feature, but it too falls well below the diffuse ISM correlation line.

\nocite{ossenkopf_henning94}
\nocite{bowey_adamson_whittet98}
\nocite{bessell_brett88}

\section{Discussion\label{discussion}}

The available data show that in the diffuse ISM, the silicate extinction correlates linearly with \ejk\ (Whittet 2003).  We present strong evidence that in dense clouds, this linear  relation breaks down  when \ejk\ exceeds 2 mag (equivalent \av\ $\sim12$ mag) (Figure~1).   Thus, the data presented also show that \tausil\ does not provide a good measure of total dust column in dense cloud environments.   It is clear from Figure 1 that \tausil\ becomes weaker per unit \ejk\ as \ejk\ increases beyond 2 mag.  The simplest explanation for this observed behavior is the effect of grain growth on the near-IR extinction. Grains of mean radius $a$ produce extinction most efficiently when  $2\pi a/\lambda \sim 1$ (Whittet 2003); for $\lambda\sim 2$ \micron, we have  $a\sim 0.3$~\micron, which is a factor of about 2 larger than the grains responsible for visual extinction in the diffuse ISM. The effect is illustrated in Fig. 2, which plots opacities from Ossenkopf \& Henning (1994) for the MRN size distribution (Mathis, Rumpl \& Nordsieck 1977, solid curve), compared with MRN subject to growth by coagulation (dashed curve) and coagulation + thin ice-mantle growth (dotted curve). Each curve is normalized relative to the opacity at 2.2~$\mu$m.  The near-IR extinction is significantly enhanced by coagulation for a given silicate opacity.  Ice mantle growth does not affect the strength of the silicate feature, but does affect the apparent shape due to blending with the H$_2$O libration feature.  This is expected since ice mantle formation adds only a negligibly thin ($< 200$ \AA) layer that will not affect near-IR extinction much \nocite{jura80,draine85ppII} (Jura 1980; Draine 1985).  Note that the shape of the 1--8~$\mu$m continuum extinction is little affected by either growth mechanism (Fig.2).  This is consistent with the observed invariance of near-IR extinction law in the Barnard 59 dense core over \av\ between 6 and 60 mag.  \nocite{romanzuniga_etal2007} (Rom{\'a}n-Z{\'u}{\~n}iga \etal\ 2007). 



The various dust models proposed to explain interstellar extinction all assume that the extinction arises from a combination of silicate and carbonaceous grains.  The main difference is in the assumed physical relationship between the silicate and carbonaceous components.  In models that assume independent silicate and carbonaceous grain components, it is primarily the carbonaceous component that is responsible for the near-IR and visual extinction and the silicate grains that dominate the extinction in the 10 \micron\ region  \nocite{mrn,draine_lee84,kim_martin94sizedist_extinction} (e.g., MRN; Draine \& Lee 1984; Kim, Martin \& Hendry 1994). In contrast, models that assume multiple independent components (small grains for the UV extinction, plus silicate cores covered with a carbonaceous mantle), the mantled silicate grains carry the extinction in the visible through the mid-IR  \nocite{li_greenberg97} (D\'esert \etal\ 1990; Li \& Greenberg 1997).  Similarly, in the models of  \nocite{mathis_whiffen89} Mathis and Whiffen (1989), the visible through mid-IR extinction is carried by a single grain component consisting of silicate and graphitic/amorphous carbon agglomerates.
Finally, \nocite{zubko_etal04} Zubko \etal\ (2004) considered a variety of dust components in their models including composite particles (containing silicate, organic refractory material, water-ice, and voids), polycyclic aromatic hydrocarbons (PAHs), bare silicate, graphite and amorphous carbon particles.  They show that the contribution to the extinction in each wavelength regime is highly model dependent.  

Regardless of the details of the adopted dust model, grain coagulation in dense clouds appreciably affects the near-IR extinction.  In dense clouds, significant grain growth is expected to result from coagulation (Jura 1980).  In the diffuse medium, grain processing is dominated by 50-100 km/s shocks. Because of the low strength of agglomerates, grain-grain collisions in even a weak interstellar shock will shatter agglomerates into their constituent ``monomers'' \nocite{jones_etal96} (Jones \etal\ 1996). Further processing, in the diffuse ISM, is then due to sputtering in shocks in the warm intercloud medium and reaccretion of gaseous silicon, magnesium, and iron atoms in an oxide mantle in diffuse clouds \nocite{savage_sembach96,tielens98} (Savage \& Sembach 1996; Tielens 1998). Thus, in diffuse clouds, these destructive processes prevent the grains from growing big enough to affect the near-IR extinction.  Such a scenario could explain the good correlation of the near-IR/visual extinction and the strength of the 9.7 \micron\ feature in the diffuse ISM and the lack of a linear correlation in dense clouds. Further laboratory studies on the coagulation behavior of mixed grain populations would be very helpful to settle this issue.

Observationally, a number of questions remain: 

\noindent$\bullet$ Is the IR extinction law invariable in all dense clouds/cores?\\
$\bullet$ At what extinction level or density does the onset of rapid dust coagulation begin?\\
$\bullet$ Does this onset vary from cloud to cloud?\\
$\bullet$ Do ice mantles affect the coagulation efficiency?

\noindent Further detailed studies that map the near-IR extinction and the 9.7 \micron\ silicate absorption feature across the extent of individual dense clouds and cores are needed and will reveal valuable information about dust properties in these regions.

\acknowledgments
This work is based [in part] on observations made with the Spitzer Space
Telescope, which is operated by the Jet Propulsion Laboratory, California
Institute of Technology under a contract with NASA. Support for this work was
provided by NASA.  The authors thank the referee, Viktor Zubko, for his helpful suggestions.

{\it Facilities:} \facility{Spitzer (IRS)}, \facility{2MASS}


\begin{thebibliography}{21}
\expandafter\ifx\csname natexlab\endcsname\relax\def\natexlab#1{#1}\fi

\bibitem[{{Becklin} {et~al.}(1978){Becklin}, {Neugebauer}, {Willner}, \&
  {Matthews}}]{becklin_etal78IV}
{Becklin}, E.~E., {Neugebauer}, G., {Willner}, S.~P., \& {Matthews}, K. 1978,
  \apj, 220, 831

\bibitem[{{Bessell} \& {Brett}(1988)}]{bessell_brett88}
{Bessell}, M.~S., \& {Brett}, J.~M. 1988, \pasp, 100, 1134

\bibitem[{{Bowey} {et~al.}(1998){Bowey}, {Adamson}, \&
  {Whittet}}]{bowey_adamson_whittet98}
{Bowey}, J.~E., {Adamson}, A.~J., \& {Whittet}, D.~C.~B. 1998, \mnras, 298, 131

\bibitem[{{Desert} {et~al.}(1990){Desert}, {Boulanger}, \&
  {Puget}}]{desert_etal90}
{Desert}, F.-X., {Boulanger}, F., \& {Puget}, J.~L. 1990, \aap, 237, 215

\bibitem[{{Draine}(1985)}]{draine85ppII}
{Draine}, B.~T. 1985, in Protostars and Planets II, ed. D.~C. {Black} \& M.~S.
  {Matthews}, 621--640

\bibitem[{{Draine} \& {Lee}(1984)}]{draine_lee84}
{Draine}, B.~T., \& {Lee}, H.~M. 1984, \apj, 285, 89

\bibitem[{{Indebetouw} {et~al.}(2005){Indebetouw}, {Mathis}, {Babler}, {Meade},
  {Watson}, {Whitney}, {Wolff}, {Wolfire}, {Cohen}, {Bania}, {Benjamin},
  {Clemens}, {Dickey}, {Jackson}, {Kobulnicky}, {Marston}, {Mercer},
  {Stauffer}, {Stolovy}, \& {Churchwell}}]{indebetouw_etal05}
{Indebetouw}, R., {Mathis}, J.~S., {Babler}, B.~L., {Meade}, M.~R., {Watson},
  C., {Whitney}, B.~A., {Wolff}, M.~J., {Wolfire}, M.~G., {Cohen}, M., {Bania},
  T.~M., {Benjamin}, R.~A., {Clemens}, D.~P., {Dickey}, J.~M., {Jackson},
  J.~M., {Kobulnicky}, H.~A., {Marston}, A.~P., {Mercer}, E.~P., {Stauffer},
  J.~R., {Stolovy}, S.~R., \& {Churchwell}, E. 2005, \apj, 619, 931

\bibitem[{{Jones} {et~al.}(1996){Jones}, {Tielens}, \&
  {Hollenbach}}]{jones_etal96}
{Jones}, A.~P., {Tielens}, A. G. G.~M., \& {Hollenbach}, D.~J. 1996, \apj, 469,
  740

\bibitem[{{Jura}(1980)}]{jura80}
{Jura}, M. 1980, \apj, 235, 63

\bibitem[{{Kim} {et~al.}(1994){Kim}, {Martin}, \&
  {Hendry}}]{kim_martin94sizedist_extinction}
{Kim}, S.-H., {Martin}, P.~G., \& {Hendry}, P.~D. 1994, \apj, 422, 164

\bibitem[{{Li} \& {Greenberg}(1997)}]{li_greenberg97}
{Li}, A., \& {Greenberg}, J.~M. 1997, \aap, 323, 566

\bibitem[{{Mathis} {et~al.}(1977){Mathis}, {Rumpl}, \& {Nordsieck}}]{mrn}
{Mathis}, J.~S., {Rumpl}, W., \& {Nordsieck}, K.~H. 1977, \apj, 217, 425

\bibitem[{{Mathis} \& {Whiffen}(1989)}]{mathis_whiffen89}
{Mathis}, J.~S., \& {Whiffen}, G. 1989, \apj, 341, 808

\bibitem[{{Ossenkopf} \& {Henning}(1994)}]{ossenkopf_henning94}
{Ossenkopf}, V., \& {Henning}, T. 1994, \aap, 291, 943

\bibitem[{{Roche} \& {Aitken}(1984)}]{roche_aitken84}
{Roche}, P.~F., \& {Aitken}, D.~K. 1984, \mnras, 208, 481

\bibitem[{{Rom{\'a}n-Z{\'u}{\~n}iga} {et~al.}(2007){Rom{\'a}n-Z{\'u}{\~n}iga},
  {Lada}, {Muench}, \& {Alves}}]{romanzuniga_etal2007}
{Rom{\'a}n-Z{\'u}{\~n}iga}, C., {Lada}, C., {Muench}, A., \& {Alves}, J. 2007,
  ArXiv e-prints, 704

\bibitem[{{Savage} \& {Sembach}(1996)}]{savage_sembach96}
{Savage}, B.~D., \& {Sembach}, K.~R. 1996, \araa, 34, 279

\bibitem[{{Tielens}(1998)}]{tielens98}
{Tielens}, A.~G.~G.~M. 1998, \apj, 499, 267

\bibitem[{{Whittet}(2003)}]{whittet_bk03}
{Whittet}, D. C.~B. 2003, Dust in the Galactic Environment, 2nd edn. (Bristol:
  Institute of Physics (IOP) Publishing)

\bibitem[{{Whittet} {et~al.}(1988){Whittet}, {Bode}, {Longmore}, {Admason},
  {McFadzean}, {Aitken}, \& {Roche}}]{whittet_etal88}
{Whittet}, D. C.~B., {Bode}, M.~F., {Longmore}, A.~J., {Admason}, A.~J.,
  {McFadzean}, A.~D., {Aitken}, D.~K., \& {Roche}, P.~F. 1988, \mnras, 233, 321

\bibitem[{{Zubko} {et~al.}(2004){Zubko}, {Dwek}, \&
  {Arendt}}]{zubko_dwek_arendt2004}
{Zubko}, V., {Dwek}, E., \& {Arendt}, R.~G. 2004, \apjs, 152, 211

\end{thebibliography}

\clearpage

\begin{deluxetable}{llccccc} 
\tabletypesize{\scriptsize}
\tablecolumns{9} 
\tablewidth{0pc} 
\tablecaption{Dense Cloud Lines of Sight} 
\tablehead{ 
\colhead{Source} &  \colhead{2MASS ID}   & \colhead{Spectral Type\tablenotemark{a}} 
& \colhead{$\tau_{9.7}$} 
& \colhead{$J-K_S$\tablenotemark{b}}
& \colhead{$E(J-K_S)$\tablenotemark{c}}
& \colhead{Spitzer PID\tablenotemark{d}} \\ 
}
\startdata 
IC~5146:Quidust 21-1 & 21472204+4734410 & G0-M4 III & 0.51 & 4.973 & 4.16 & 3320 \\
IC~5146:Quidust 21-2 & 21463943+4733014 & G0-M4 III &  0.55 & 2.733 & 1.92 & 3320 \\
IC~5146:Quidust 21-3 & 21475842+4737164 & G0-M4 III & 0.50 & 2.619 & 1.81 & 3320 \\
IC~5146:Quidust 21-4 & 21450774+4731151 & G0-M4 III & 0.34 & 2.185 & 1.38 & 3320 \\
IC~5146:Quidust 21-5 & 21444787+4732574 & G0-M4 III & 0.33 & 2.040 &1.23 & 3320 \\
IC~5146:Quidust 21-6 & 21461164+4734542 & G0-M4 III & 0.60 & 4.337 & 3.53 & 3320 \\
IC~5146:Quidust 22-1 & 21443293+4734569 & G0-M4 III & 0.55 & 3.780 & 2.97 & 3320 \\
IC~5146:Quidust 22-3 & 21473989+4735485 & G0-M4 III & 0.24 & 1.197 & 0.39 & 3320 \\
IC~5146:Quidust 23-1 & 21473509+4737164 & G0-M4 III & 0.50 & 1.917 & 1.11 & 3320 \\
IC~5146:Quidust 23-2 & 21472220+4738045 & G0-M4 III & 0.60 & 2.474 & 1.66 & 3320 \\
Taurus:Elias 3   & 04232455+2500084 & K2 III & 0.40 & 2.516 & 1.76 & 172 \\
Taurus:Elias 9   & 04321153+2433380 & K2 III & 0.21\tablenotemark{e} & 2.109 & 0.93 & \nodata \\
Taurus:Elias 13 & 04332592+2615334 & K2 III & 0.54\tablenotemark{e} & 3.005 & 2.26 & \nodata \\
Taurus:Elias 16 & 04393886+2611266 & K1 III & 0.75\tablenotemark{f} & 5.443 & 4.76 & \nodata \\
Taurus:HD 29647 & 04410804+2559340 & B8 III & 0.24\tablenotemark{e} & 0.597 & 0.16 & \nodata \\
Barnard 59:          & 17111538-2727144 & G0-M4 III      &  0.85 &  $>6.062$ & 5.58 & 20604 \\
Barnard 59:          & 17112005-2727131 & G0-M4 III      & 0.80  & $>4.409$  &  8.86 & 20604 \\
Barnard 68:Quidust 18-1 & 17224500-2348532 & G0-M4 III & 0.2 & 1.637 & 0.83 & 3320 \\
Barnard 68:Quidust 19-1 & 17224483-2349049 & G0-M4 III & 0.2 & 1.720 &0.93 & 3320 \\
Barnard 68:Quidust 20-1 & 17224407-2349167 & G0-M4 III & 0.45 & 1.959 & 1.15 & 3320 \\
Barnard 68:Velucores 1-1 & 17223790-2348514 & G0-M4 III & 0.5 & 2.938 & 2.13 & 3290 \\
Barnard 68:Velucores 1-2 & 17224511-2348394 & G0-M4 III & 0.3 & 1.440 & 0.63 & 3290 \\
Barnard 68:Velucores 1-3 & 17224027-2348555 & G0-M4 III & 0.4 & 3.276 & 2.47 & 3290 \\
Barnard 68:Velucores 1-4 & 17224159-2350261 & G0-M4 III & 0.3 & 4.044 & 3.23 & 3290 \\
Chameleon I: Quidust 2-1 & 11024279-7802259 & G0-M4 III & 0.4 & 1.701 & 0.89 & 3290 \\
Chameleon I: Quidust 2-2 & 11055453-7735122 & G0-M4 III & 0.6 & 3.375 & 2.57 & 3290 \\
Chameleon I: Quidust 3-1 & 11054176-7748023 & G0-M4 III & 0.5 & 2.361 & 1.55 & 3290 \\
Serpens: CK 2                  & 18300061+0115201 & mid K-early M III & 1.8 & $>8.676$ & 9.45 & 172 \\
Serpens: SVS76 Ser 9     & 18294508+0118469 & G0-M4 III & 0.8 & 4.326  & 3.52 & 172 \\
Serpens: SSTc2d182852.7+02824 & 18285266+0028242 & G0-M4 III & 1.2 & 4.136 & 3.33 & 172 \\

\enddata
\tablenotetext{a}{Where spectral type range is given, it is based on placement in $J-H$ versus $H-K_s$ diagram, except for CK2.  CK2 spectral type from Chiar et al. 1994.  Taurus spectral types from Whittet et al. 1988 and references therein.}
\tablenotetext{b}{2MASS photometry.  Where lower limit is given, the source is not detected in the 2MASS $J$ band due to its faintness.}
\tablenotetext{c}{Computed using spectral types listed and associated intrinsic colors from Bessell \& Brett 1988}
\tablenotetext{d}{Spitzer program ID for silicate spectra, where applicable.  PID 3320, this work; PID 172, PI Evans, c2d Legacy; PID 3290, PI Langer.}
\tablenotetext{e}{Silicate measurement from Whittet et al. 1988}
\tablenotetext{f}{Silicate measurement from Bowey, Adamson \& Whittet 1998}

\end{deluxetable}

\clearpage

\begin{figure}
\includegraphics[angle=90,scale=0.7]{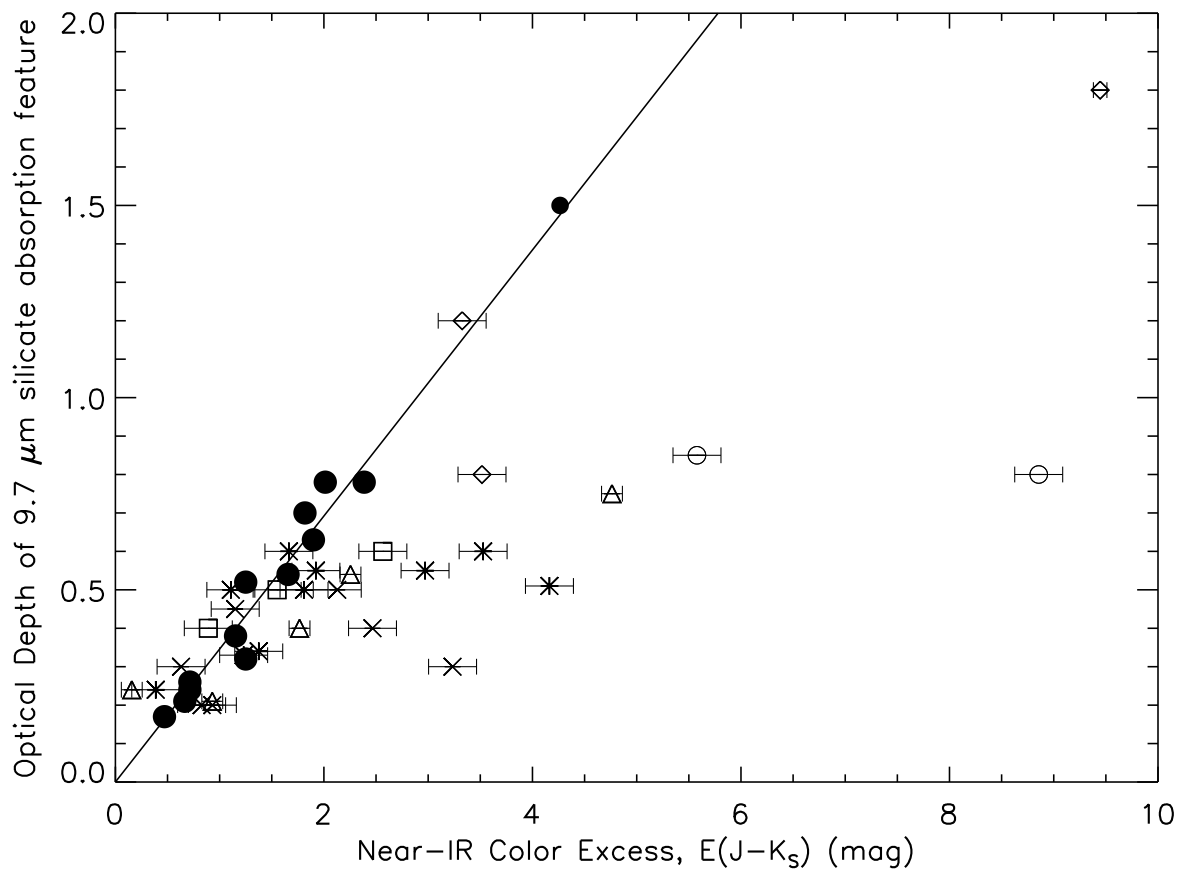}
\caption{Plot of the relationship between the optical depth of the 9.7 \micron\ silicate absorption feature versus the near-IR color excess, \ejk, for Taurus (triangles), IC 5146 (asterisks), Serpens (diamonds), Chameleon I (squares), Barnard 68 (X's), Barnard 59 (open circles).  Diffuse ISM lines of sight are represented by large filled circles (Whittet 2003, and references therein) and a small filled circle (Chiar, unpublished Spitzer data).  The diagonal line represents the ``diffuse ISM'' correlation.  Error bars indicate spectral type uncertainties in the calculation of \ejk.
\label{fig:silav}}
\end{figure}

\clearpage

\begin{figure}
\includegraphics[angle=0,scale=0.7]{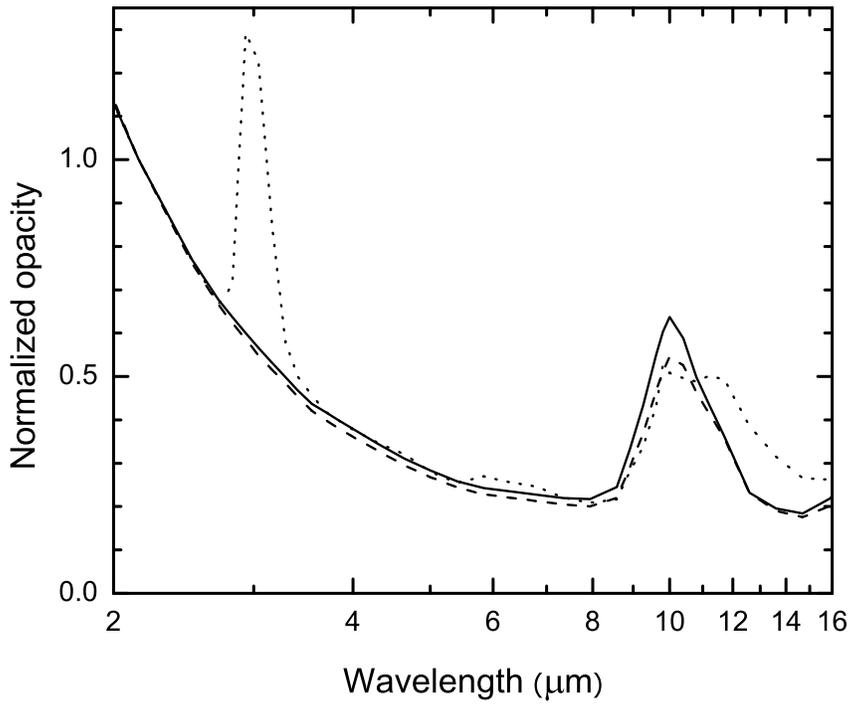}
\caption{Plot of dust opacities vs.\ wavelength. The solid curve is for the original MRN grain model applicable to the diffuse ISM. The dashed curve shows the effect of growth by coagulation after $10^5$ years at a particle density $n=10^6$~cm$^{-3}$. The dotted curve shows the effect of thin ice-mantle growth in addition to coagulation. Each curve is normalized relative to the opacity at 2.2~$\mu$m. Data are from Table~1 of Ossenkopf \& Henning (1994).
\label{fig:coag}}
\end{figure}

\end{document}